\documentstyle[twoside,fleqn,espcrc2,epsf]{article}

\newcommand{\Dslash}{$D$\kern-0.6em \hbox{/}}

\newcommand{\la}{\raise.16ex\hbox{$\langle$}}
\newcommand{\ra}{\raise.16ex\hbox{$\rangle$}}
\newcommand{\be}{\begin{equation}}
\newcommand{\ee}{\end{equation}}
\newcommand{\bea}{\begin{eqnarray}}
\newcommand{\eea}{\end{eqnarray}}

\title{UV-filtered fermionic Monte Carlo}

\author{
Ph. de Forcrand\address{SCSC, ETH-Z\"urich, CH-8092 Z\"urich, Switzerland \\
({\bf forcrand@scsc.ethz.ch})} }

\begin{document}

\begin{abstract}
The short-range modes of the fermionic determinant can be absorbed in
the gauge action using the loop expansion. The coefficients of this
expansion and the zeroes of the polynomial approximating the remainder
can be optimized by a simple, practical method. When the multiboson
approach is used, this optimization results in a faster simulation with fewer
auxiliary fields.
\end{abstract}
\maketitle

Dynamical fermion simulations are orders of magnitude slower than 
quenched ones. With Hybrid Monte Carlo (HMC), this remains true
also for heavy dynamical quarks: even in the quenched regime,
HMC is ${\cal O}(100)$ less efficient than a local Monte Carlo update
scheme \cite{Sharpe}. This poor behaviour may explain why HMC computing
requirements
grow rather slowly as the quark mass is decreased. It clearly indicates
room for improvement, which hopefully will persist down to light quarks.

The fermion determinant can be expanded in loops:
\be
{\rm det}({\bf 1} - \kappa M) = e^{{\rm Tr~Log}({\bf 1} - \kappa M)}
= e^{- \sum_l \frac{\kappa^l}{l} {\rm Tr}~M^l} 
\ee
using Wilson fermion notation. The first nonzero term ($l=4$) gives
a shift $\Delta \beta$ of the gauge coupling. It is remarkable that the
bulk of the fermionic effects can be reabsorbed into this simple shift,
which can be computed perturbatively down to rather small quark masses
\cite{DeGrand}. Unfortunately, neither HMC nor the alternative MultiBoson
(MB) method makes use of this observation. 
Effective loop actions (\cite{Sexton}) $S_{{\rm eff}}$ cannot be used to
guide an exact algorithm, because 
$\langle (S_{{\rm exact}} - S_{{\rm eff}})^2 \rangle \propto Volume$, 
which causes an exponentially small $e^{-Volume}$ Metropolis acceptance.
Instead, I propose to make use of the identity
$e^{-{\rm Tr}~A}\times{\rm det}~e^{A} = 1$
to rewrite the fermion determinant as
\bea
\lefteqn{ {\rm det}({\bf 1} - \kappa M) \equiv } \\
\label{det}
& e^{- \sum_j a_j {\rm Tr} M^j} \times
{\rm det}\left( ({\bf 1} - \kappa M) ~~e^{+ \sum_j a_j M^j} \right) \nonumber
\eea
The number of nonzero coefficients $a_j$ and their values are subject to optimization.
Note that the first four terms cause no overhead. 
The new operator whose determinant is to be taken will hopefully be
easier to sample than the original one. In particular, for very heavy quarks 
the coefficients $a_j$ should tend to the Taylor values
$\kappa^j / j$, leaving only tiny higher-order terms in the determinant.

Both HMC and MB sample the determinant ${\rm det}A$ by building a 
polynomial $P_n(A)$ which approximates $A^{-1}$. Gains in efficiency
come from a better approximation (smaller degree or smaller error).
To illustrate the advantage of using Eq.(2), in Fig.1 I show
the relative error of the polynomial approximation for a real variable
$x \in [0.01,1]$. The dotted line corresponds to the Chebyshev approximation
used in the MB method; the solid line is the error achievable by
including two terms $a_1$ and $a_2$ in the ``preconditioner''
$e^{- \sum_j a_j x^j}$, with a polynomial of the same degree $n=20$.
Fig.2 shows the zeroes of the two polynomials in the complex plane.
While the zeroes of the Chebyshev polynomial are distributed uniformly
over an ellipse covering the entire approximation interval, the other
zeroes are much more concentrated near the origin. 
This is the essence of the benefits of the ``filter'' $e^{- \sum_j a_j x^j}$:
large eigenvalues (corresponding to UV modes of the Dirac operator) are
effectively removed from the determinant, which frees the zeroes of the polynomial
to concentrate near small eigenvalues corresponding to IR modes and
genuine long-range fermionic interactions.

\begin{figure}[t]
\begin{center}
\epsfxsize=6.5cm 
\epsfysize=6.1cm 
\epsffile[57 195 571 669]{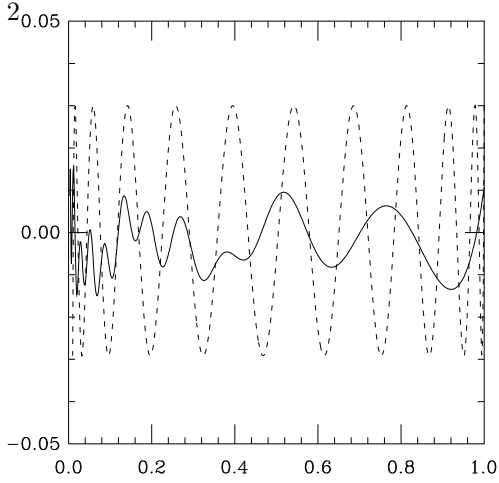}
\end{center}
\vspace{-2.8cm}
\caption{Error of the Chebyshev (dotted line) and 
UV-filtered (solid line) approximations for a polynomial of degree 20
over the interval $[0.01,1]$.}
\vspace{-0.2cm}
\end{figure}

Choosing the coefficients $\{a_j\}$ and the zeroes $\{z_k\}$ of the
polynomial approximation seems like a stiff nonlinear optimization problem,
with the $a_j$ appearing in the exponent. Even setting $a_j=0 ~\forall j$,
the choice of the zeroes $\{z_k\}$ has been the object of several studies
\cite{Borici,Montvay}. The problem is how to fix the parameters of
$W \equiv \prod_k^n ({\bf 1} - \kappa M - z_k {\bf 1}) \cdot
({\bf 1} - \kappa M) \cdot e^{\sum_{j=0}^{m-1} a_j M^j} $
such that ${\rm det}W \approx 1$. Rewriting the determinant as an
average over Gaussian random vectors $\eta$
\be
{\rm det}^{-2} W = \frac{\int d\eta^\dagger d\eta e^{-\eta^\dagger W^\dagger W 
\eta}}
{\int d\eta^\dagger d\eta e^{-\eta^\dagger \eta}}
= \langle e^{- |W\eta|^2 + |\eta|^2} \rangle_{\eta}
\ee
we see that a sufficient condition for ${\rm det}W \approx 1$ is
$\parallel W\eta - \eta \parallel^2 \approx 0$.
This optimization problem is easy to solve, using the following steps: \\
1. Draw one (or more) Gaussian vectors $\eta$; \\
2. Assign values to the $\{a_j\}$; compute
$~ \psi \equiv ({\bf 1} - \kappa M) ~ e^{\sum_{j=0}^{m-1} a_j M^j} ~ \eta$; \\
3. Construct the polynomial 
$Q_n(M) \equiv \prod_k^n ({\bf 1} - \kappa M - z_k {\bf 1})$ 
which minimizes $c \equiv \parallel Q_n(M) \psi - \eta \parallel^2$; \\
4. Compute $\vec{\nabla}_a \equiv \left\{ \frac{\partial c}{\partial a_j} \right \}$
as a by-product; if $\parallel \vec{\nabla}_a \parallel > \epsilon$, return to (2).
 \\
Step 3 is a straightforward quadratic minimization similar to GMRES, 
while the loop 2--4 uses Newton's
method to solve the nonlinear minimization in the $a_j$.
In principle, some subtle averaging over gauge fields should also be performed;
in practice, different equilibrium gauge fields yield similar results.

This optimization method is simple and practical. It requires no {\em a priori}
knowledge of the Dirac spectrum. It is applicable to all cases where a
determinant is involved: Wilson or staggered fermions, SUSY, condensed matter. 
When applied to the standard MB method ($\{a_j\}=0$, Hermitian or non-Hermitian
Wilson Dirac operator), it reveals that the usual Chebyshev approximation is not
optimal. 

In an actual MC simulation $\{U_{{\rm old}}\} \rightarrow \{U_{{\rm new}}\}$,
the Metropolis acceptance probability is 
$min(1,\langle \frac{e^{- |W_{{\rm old}}\eta|^2 + |\eta|^2}}
{e^{- |W_{{\rm new}}\eta|^2 + |\eta|^2}} \rangle_{\eta})$.
This can be estimated as ${\rm erfc}(c \parallel W_{{\rm old}} \eta - \eta \parallel)$
with $c \sim {\cal O}(1)$ as a by-product of the above optimization.
In this way, the benefits of adding more terms to the preconditioner
(especially 6-link and larger loops) can be assessed {\em without} having to
program the actual Monte Carlo update.

\begin{figure}[t]
\begin{center}
\epsfxsize=6.5cm 
\epsfysize=6.1cm 
\epsffile[69 195 558 662]{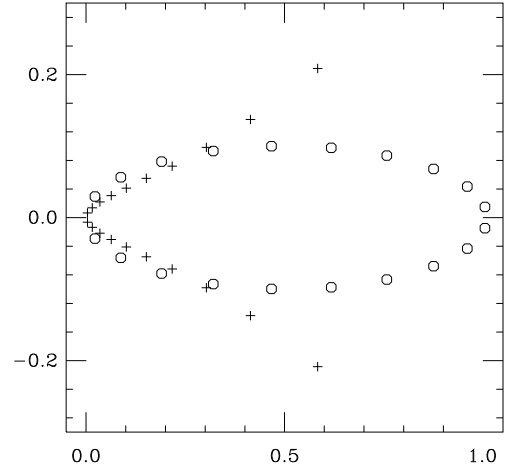}
\end{center}
\vspace{-2.8cm}
\caption{Zeroes of the Chebyshev ($\circ$) and UV-filtered (+) polynomials
in the complex plane.}
\end{figure}

Tests of the dynamical behaviour of this UV-filtered MB algorithm
have been performed for moderate quark masses: $\beta=5.3, \kappa=0.158$,
with two flavors of Wilson fermions, on an $8^4$ lattice. Table I compares 
the efficiency of HMC, the non-Hermitian MB \cite{Galli}, and the present
method. All the programs used even-odd preconditioning. The HMC program
incorporates multiple step-size integration \cite{SW} and low-accuracy
solution during the trajectory \cite{TT}, and uses the 
BiCG$\gamma_5$ solver \cite{Melbourne}. The number $n$ of fields used by the
non-Hermitian MB is consistent with the number of iterations of the HMC
solver. In contrast, the UV-filtered version uses $\sim 3$ times
fewer fields. This solves the memory bottleneck of the MB approach
and dramatically reduces the work per independent configuration, which
grows like $\sim n^2$ \cite{Melbourne}. Fig.3 shows the autocorrelation
function of the plaquette for HMC and the UV-filtered MB.
For heavier quarks the superiority of the latter would be even greater.
For lighter quarks the advantage is less pronounced, and details of
the implementation and the tuning (eg., over-relaxation of the auxiliary
fields) become more relevant.

\begin{table}
\begin{center}
\begin{tabular}{|c|c|c|c|}
\hline
\hline
$\beta=5.3, \kappa=0.158$   & HMC & MB & this work \\
\hline
$\Delta \beta$ & 0 & 0 & 0.166 \\
\hline
deg. of polynom. & 26/65 & 20 & 7 \\
\hline
$\tau_{int}(\Box)$ in $\Dslash \times \vec{v}$   & $\sim 16000$ & $\sim 38000$ & $\sim 3200$ \\
\hline
\hline
\end{tabular}
\end{center}
\caption{Comparison of three exact algorithms: Hybrid Monte Carlo, MultiBoson,
and UV-filtered MultiBoson. $\Delta \beta$ is the shift in the gauge coupling
induced by UV-filtering. In the case of HMC, the degree of the polynomial is
the average number of iterations required by the BiCG$\gamma_5$ solver,
during (26) and at the ends (65) of each trajectory. The 
integrated autocorrelation time for the plaquette is measured 
in units of multiplications by \Dslash.}
\vspace{-0.5cm}
\end{table}

\begin{figure}[t]
\begin{center}
\epsfxsize=7.0cm 
\epsfysize=6.5cm 
\epsffile[40 60  318 302]{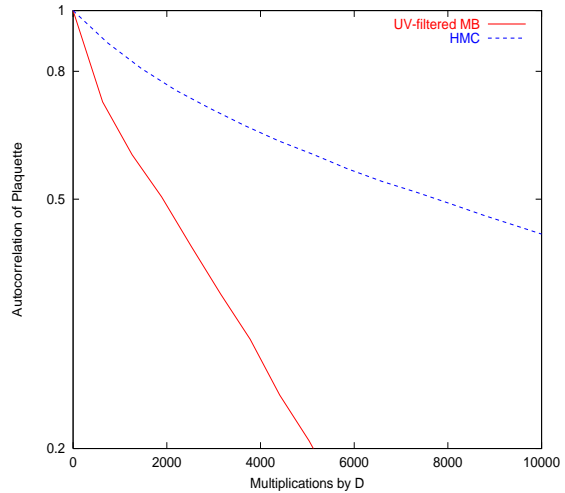}
\end{center}
\vspace{-1.5cm}
\caption{Autocorrelation of the plaquette as a function of multiplications
by \Dslash $~$ for HMC and UV-filtered MB.}
\vspace{-0.5cm}
\end{figure}

\begin{figure}[b]
\vspace{-4.0cm}
\begin{center}
\epsfxsize=6.5cm 
\epsfysize=6.5cm 
\epsfclipoff
\epsffile[39 195 508 662]{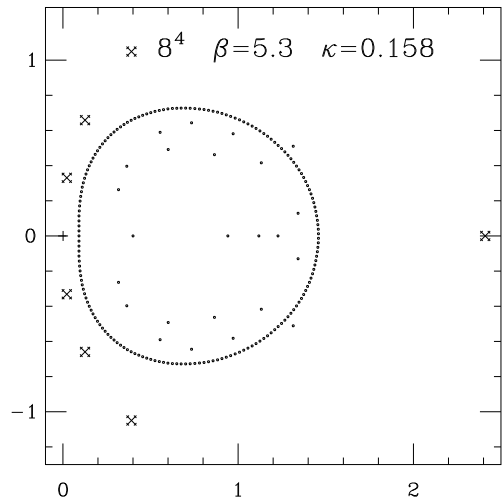}
\end{center}
\vspace{-2.8cm}
\caption{Complex spectrum of $({\bf 1} - \kappa^2 M^2)$ estimated
from the tridiagonal matrix generated by the BiCG$\gamma_5$ solver,
and zeroes of the polynomial used by the UV-filtered MB.
Only one of the seven zeroes is devoted to controlling UV modes, while the
other six are dedicated to the IR ones.}
\end{figure}

\end{document}